\documentclass{article}

\usepackage{amsfonts,amssymb,amsmath, amsthm, color, a4wide}
\usepackage{mathtools}
\usepackage{mathrsfs}
\usepackage{hyperref}
\usepackage{graphicx}
\usepackage{epstopdf}
\usepackage{bbm}
\usepackage{multirow}
\usepackage[margin=2cm]{geometry}
\usepackage[font=small,labelfont=bf]{caption}

\usepackage[round]{natbib}
\setcitestyle{notesep={: }}
\setcitestyle{aysep={,}}

\numberwithin{equation}{section}

\usepackage{authblk}

\renewenvironment{abstract}{%
	\vspace{6pt}%
	\begin{center}%
		\begin{minipage}{320pt}%
			\small%
			\begin{center}%
				\textbf{Abstract}%
			\end{center}%
		}{\end{minipage}\end{center}}
\newcommand{\keywords}[1]{%
	\begin{center}%
		\begin{minipage}{320pt}%
			\textit{Key words:}~{#1}
		\end{minipage}%
	\end{center}%
}

\title{Determinantal shot noise Cox processes}

\author[1]{Jesper M\o ller}
\author[1]{Ninna Vihrs}
\affil[1]{Department of Mathematical Sciences, Aalborg University}

\begin{document}

\maketitle

\begin{abstract}
{We present a new class of cluster point process models, which we call determinantal shot noise Cox processes (DSNCP), with repulsion between cluster centres. They are the special case of generalized shot noise Cox processes where the cluster centres are determinantal point processes. We establish various moment results and describe how these can be used to easily estimate unknown parameters in two particularly tractable cases, namely when the offspring density is isotropic Gaussian and the kernel of the determinantal point process of cluster centres is Gaussian or like in a scaled Ginibre point process. Through a simulation study and the analysis of a real point pattern data set we see that when modelling clustered point patterns, a much lower intensity of cluster centres may be needed in DSNCP models as compared to shot noise Cox processes.}
\end{abstract}

\keywords{
Gaussian determinantal point process, generalised shot noise Cox process, Ginibre point process, global envelope test, minimum contrast estimation, Thomas process} 

\section{Introduction}\label{intro}

This paper studies a cluster point process model defined as follows. Let $Y$ be a simple locally finite point process defined on the $d$-dimensional Euclidean space $\mathbb R^d$; we can view $Y$ as a random subset of $\mathbb R^d$ (for background material on spatial point processes, see \cite{MW2004}). Assume $Y$ is stationary, that is, its distribution is invariant under translations in $\mathbb R^d$.   Conditioned on $Y$, let $X$ be a Poisson process on $\mathbb R^d$ with intensity function 
\begin{equation}\label{e:rhogivenY}
\rho(x\mid Y)=\gamma\sum_{y\in Y}k_\alpha(x-y),\quad x\in\mathbb R^d,
\end{equation}
where $\gamma>0$ and $\alpha$ are parameters and $k_\alpha$ is a probability density function (pdf) on $\mathbb R^d$; in our specific models $\alpha$ will play the role of a band width (a positive scale parameter). We can identify $X$ by a cluster process $\cup_{y\in Y}X_y$ where conditioned on $Y$, the clusters $X_y$ are independent finite Poisson processes on $\mathbb R^d$ and $X_y$ has intensity function $\rho_y(x)= \gamma k_\alpha(x-y)$ (depending on the `offspring' density $k_\alpha$ relative to the cluster center $y$). 

In the special case where $Y$ is a stationary Poisson process, $X$ is a shot noise Cox process (SNCP), see \cite{moeller:02}. Then there may be a large amount of overlap between the clusters unless the intensity of $Y$ is small as compared to the band width $\alpha$. In this paper, we will instead be interested in repulsive point process models for $Y$. This may be an advantage since the repulsiveness of $Y$ implies less overlap of clusters. Thereby it may be easier to apply statistical methods for cluster detection, and when modelling clustered point pattern data sets a much lower intensity of $Y$ may be needed as compared to the case of a SNCP. The idea of using a repulsive point process $Y$ is not new, where \cite{Lies02} suggested to use a Markov point process. However, we are in particular interested in the case where $Y$ is a stationary determinantal point process (DPP) in which case we call $X$ a {\em{determinantal shot noise Cox process (DSNCP)}}. Briefly, a DPP is a model with repulsion at all scales, cf.\ \cite{LMR15}, \cite{MReilly21}, and the references therein.  There are several advantages of using a DPP for $Y$: In contrast to a Markov point process, there is no need of MCMC when simulating a DPP, and as we shall see $Y$ and hence $X$ possess nice moment results, which can be used for estimation. 

The cluster point process $X$ given by \eqref{e:rhogivenY} is a special case of a stationary generalized shot noise Cox process (GSNCP), see \cite{moeller:torrisi:2005}, and it may be extended as follows. Suppose $X$ conditioned on both $Y$ and positive random variables $\{\Gamma_y\}_{y\in \mathbb R^d}$ and $\{A_y\}_{y\in \mathbb R^d}$ is a Poisson process with intensity function
\[\rho(x\mid Y,\{\Gamma_y\}_{y\in \mathbb R^d},\{A_y\}_{y\in \mathbb R^d})=\sum_{y\in Y}\Gamma_y k_{A_y}(x-y),\quad x\in\mathbb R^d,\]
where $k_{A_y}$ is a pdf on $\mathbb R^d$. In addition, assume that $\{\Gamma_y\}_{y\in \mathbb R^d}$, $\{A_y\}_{y\in \mathbb R^d}$, and $Y$ are mutually independent, the $\Gamma_y$ are independent identically distributed with mean $\gamma$ and has finite variance, and the $A_y$ are independent identically distributed. Then $X$ is still a stationary GSNCP and if also $Y$ is a stationary DPP we may call $X$ a DGSNCP. In fact, all results and statistical methods used in this paper will apply for the DGSNCP when $k_\alpha$ is replaced by $\mathrm E k_{A_y}$ in all expressions to follow. The DGSNCP may most naturally be treated in a MCMC Bayesian setting using a similar approach as in \cite{Beraha22} and the references therein.

In the present paper, we study and exploit for statistical inference the nice moment properties for various DSNCP models as follows. In Section~\ref{sec:DSNCP} we describe further what it means that $Y$ is a DPP and present two specific cases of DSNCP models where we let $k_\alpha$ be an isotropic Gaussian density as in the Thomas process \citep{Thomas}, which is the most popular example of a SNCP. Section~\ref{s:mom} considers general results for so-called pair correlation and $K$-functions for first the GSNCP model and second the DSNCP model. Section~\ref{s:appl} discusses how the results in Section~\ref{s:mom} may be used when fitting a parametric DSNCP model in a frequentist setting, and we illustrate this on a real data example. In Section~\ref{sec:GETsimstudy} we investigate the ability to distinguish between DSNCPs and Thomas processes through a simulation study. Finally, in Section~\ref{sec:conclusion} we summarize our results.

All statistical analyses were made with \texttt{R} \citep{R}, and all plots were made using the package \texttt{ggplot2} \citep{ggplot}.

\section{Determinantal shot noise Cox process models}\label{sec:DSNCP}

In this section we consider the DSNCP model for $X$ and suggest some specific models. In brief, the DPP $Y$ is specified by a so-called kernel which is usually assumed to be a complex covariance function $c(u,v)$ defined for all $u,v\in\mathbb R^d$; for details, see Appendix~A. We assume $Y$ is a stationary DPP with intensity $\rho_Y>0$, meaning two things: First, if $A\subset\mathbb R^d$ is a bounded Borel set, $Y(A)$ denotes the cardinality of $Y\cap A$, and $|A|=\int_A\,\mathrm du$ is the Lebesgue measure of $A$,
then $\mathrm EY(A)=\rho_Y|A|<\infty$. Second, $|c(u,v)|=|c(u-v,0)|$ for all $u,v\in\mathbb R^d$ where $|s|$ denotes the modulus of a complex number $s$. We denote the corresponding complex correlation function by
$r(u,v)=c(u,v)/\rho_Y$ and assume it depends on a correlation/scale parameter $\beta>0$ so that $r=r_\beta$ with
\begin{equation}\label{e:rbeta}
r_\beta(u,v)=r_1(u/\beta,v/\beta).
\end{equation}   

 The correlation parameter $\beta$ cannot vary independently of $\rho_Y$ since there is a trade-off between intensity and repulsiveness in order to secure that a DPP model is well defined \citep{LMR15}. For instance, for many DPP models $r_\beta$ is real, continuous, and stationary, that is, $r_\beta(u,v)=r_{\beta,\mbox{st}}(u-v)$ where $r_{\beta,\mbox{st}}:\mathbb R^d\to[-1,1]$ is a continuous, symmetric, and positive semi-definite function with $r_{\beta,\mbox{st}}(0)=1$. Then, if $r_{\beta,\mbox{st}}$ is square integrable and has Fourier transform $\varphi_\beta(u)=\int r_{\beta,\mbox{st}}(v)\cos(2\pi u\cdot v)\,\mathrm dv$ where $\cdot$ is the usual inner product, the DPP is only well-defined for $\rho_Y\sup\varphi_\beta\le1$, cf.\ \cite{LMR15}. In case of \eqref{e:rbeta}, this existence condition of the DPP means that $0<\beta\le1/(\rho_Y^{1/d}\sup\varphi_1)$, where for a fixed value of $\rho_Y$, most repulsiveness is obtained when $\beta=1/(\rho_Y^{1/d}\sup\varphi_1)$. 

Consider the special case where $Y$ is a jinc-like DPP, that is, $d=2$ and $r(u,v)=J_1(2\sqrt\pi\|u-v\|)/(\sqrt\pi\|u-v\|)$ where $J_1$ is the first order Bessel function of the first kind and $\|\cdot\|$ denotes usual distance. So, the distribution of $Y$ depends only on the intensity, and $Y$ is a most repulsive DPP in the sense of \cite{LMR15}, see also \cite{biscio16} and \cite{MReilly21}. \cite{Andreasmfl} used this special case of a DSNCP model in a situation where a realization of $X$ but not $Y$ was observed within a bounded region. They estimated $\rho_Y$ with a minimum contrast procedure based on the pair correlation function (pcf) given by \eqref{e:gXDPP} in Section~\ref{s:mom}, where the pcf 
had to be approximated by numerical methods. Instead we consider more tractable cases, as we shall see in Section~\ref{s:mom}. 
   
Note that $X$ is stationary with intensity $\rho_X=\gamma\rho_Y$. We will consider two specific DSNCP models of $X$ where we let $k_\alpha$ be the pdf of $N_d(0,\alpha^2I)$, the zero-mean isotropic $d$-dimensional normal distribution, and $Y$ is given by one of the following two DPPs.
\begin{enumerate}
\item If $r_\beta(u,v)=\exp\left(-\|(u-v)/\beta\|^2\right)$ is the Gaussian correlation function, 
then $Y$ is a {\em{Gaussian DPP}} \citep{LMR15}.
\item If $d=2$ and we identify $\mathbb R^2$ with the complex plane $\mathbb C$, and if $r_\beta(u,v)=\exp((u\overline {v}-|u|^2/2-|v|^2/2)/\beta^2)$ 
where $\overline {v}$ and $|v|$ denote the complex conjugate and the modulus of the complex number $v$, then $Y$ is a {\em{scaled Ginibre point process}} (\citet{Deng,Miyoshi}; $Y$ is the standard Ginibre point process \citep{Ginibre} if $\rho_Y=1/\pi$ and $\beta=1$). 
\end{enumerate}
Both these DPP models are well defined if and only if $0<\beta\le 1/(\rho_Y^{1/d}\sqrt\pi)$ \citep{LMR15}. For a fixed value of $\rho_Y$, $Y$ becomes in both cases less and less repulsive as $\beta$ decreases from $1/(\rho_Y^{1/d}\sqrt\pi)$ to $0$, where in the limit $Y$ is a stationary Poisson process and $X$ is a Thomas process. Therefore, we call $X$ a {\em{Gaussian-DPP-Thomas process}} in the first case and a {\em{Ginibre-DPP-Thomas process}} in the second case. In both cases, $Y$ and hence also $X$ are stationary and isotropic,
 although $r_\beta$ is only stationary when it is the Gaussian correlation function (Appendix~B1 verifies that the distribution of a scaled Ginibre point process $Y$ is invariant under isometries).

\begin{figure}[ht!]
\centering
\includegraphics[scale=1]{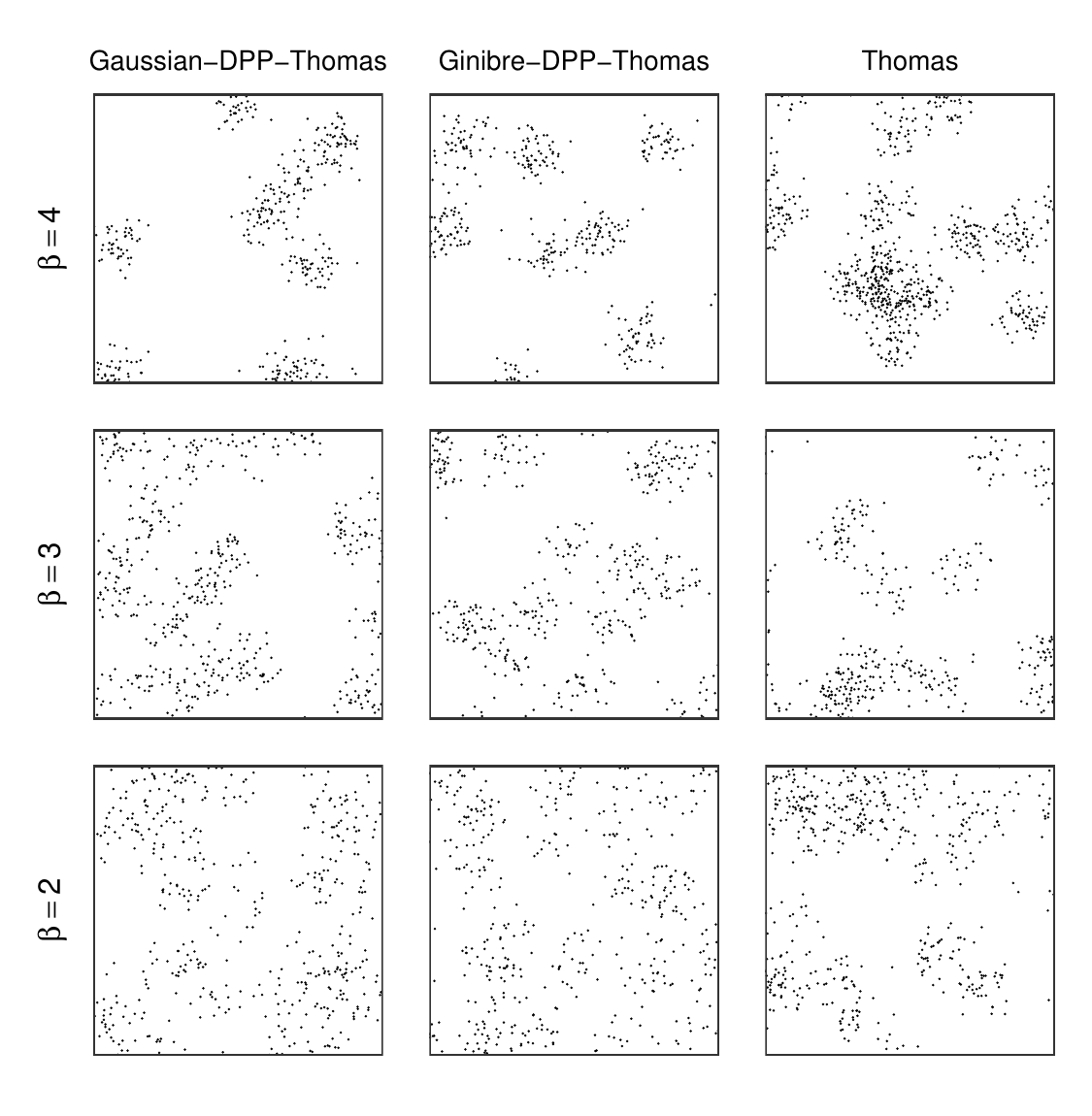}
\caption{Simulations of Gaussian-DPP-Thomas processes, Ginibre-DPP-Thomas processes, and Thomas processes (as stated at the top of each column) within a square with side lengths 20 when $\alpha=1$, $\rho_X=1$, $\beta = 2,3,4$ (stated to the left of each row), and $\rho_Y=1/(\pi\beta^2)$ in all processes. Note that $\beta$ is not a parameter of the Thomas process and is thus only used to calculate $\rho_Y$ in this case.}
\label{fig:sims}
\end{figure}

The two first columns of plots in Figure~\ref{fig:sims} show simulated realizations of a Ginibre-DPP-Thomas process and a Gaussian-DPP-Thomas process within a $20\times20$ square region, where $\alpha=1$,
 $\rho_X=1$ (so we expect to see about 400 points in each simulated point pattern), and in the three rows of plots we have $\beta=2,3,4$ (from bottom to top). In each case, $\rho_Y=1/(\pi\beta^2)$ is as large as possible. For comparison, the third column of plots in Figure~\ref{fig:sims} 
shows simulations of Thomas processes with the same values of $(\rho_X,\rho_Y)$ as for the two first columns of plots. So, in each row the three processes have the same expected number of clusters, the same expected cluster sizes, and the same offspring density. For all processes, as $\beta$ increases (that is, $\rho_Y$ decreases and $\gamma$ increases) we see  that the point patterns look more clustered, since we get less and less cluster centres but larger and larger clusters. We also see that the eye detects less diffuse clusters in the DPP-Thomas processes compared to the Thomas processes, which is in agreement with the fact that cluster centres are repulsive in DPP-Thomas processes whereas they are completely random in Thomas processes. From Figure~\ref{fig:sims} it can be difficult to make any conclusions about the differences between Gaussian- and Ginibre-DPP-Thomas processes, but we will make further comparisons between these in Sections~\ref{s:mom}--\ref{s:appl}.

\section{Pair correlation and $K$-functions}\label{s:mom}

\subsection{The general setting of stationary GSNCPs}

Consider again the general setting in Section~1 of a stationary GSNCP.
Henceforth we assume the stationary point process $Y$ has pair correlation function (pcf) $g_Y$. This means that if $A,B\subset\mathbb R^d$ are disjoint bounded Borel sets, then
\[\mathrm E[Y(A)Y(B)]=\rho_Y^2\int_A\int_B g_{Y}(u,v)\,\mathrm du\,\mathrm dv<\infty.\]
By stationarity, $g_Y(u,v)=g_{Y,\mbox{st}}(u-v)$ depends only on the lag $u-v$ almost surely (with respect to Lebesgue measure) and for ease of presentation we can assume this is the case for all $u,v\in\mathbb R^d$.

The stationary GSNCP $X$ given by \eqref{e:rhogivenY} has 
intensity
 $\rho_X=\gamma\rho_Y$ and a stationary pcf $g_X(u,v)=g_{X,\mbox{st}}(x)$ where $x=u-v$ and
\begin{equation}\label{e:gX}
g_{X,\mbox{st}}(x)=k_\alpha*{\tilde k_\alpha}*g_{Y,\mbox{st}}(x)+k_\alpha*{\tilde k_\alpha}(x)/\rho_Y,\quad x\in\mathbb R^d,
\end{equation}
where $*$ denotes convolution and ${\tilde k_\alpha}(x)=k_\alpha(-x)$ is the reflection of $k_\alpha$, cf.\ \cite{moeller:torrisi:2005}. Thus, $g_X$ decreases as $\rho_Y$ increases; this makes good sense since $\rho_Y$ is the intensity of clusters and the first term in \eqref{e:gX} corresponds to pairs of points from different clusters whilst the second term is due to pairs of points within a cluster. Furthermore, from \eqref{e:gX} we obtain Ripley's $K$-function \citep{ripley:76,ripley:77}  
 \[K(r)=\int_{\|x\|\le r}g_{X,\mbox{st}}(x)\,\mathrm dx,\quad r>0,\]
which will be used in Section~\ref{s:appl} for parameter estimation. 
  
In the special case where $Y$ is a stationary Poisson process (i.e., $X$ is a SNCP), we have $g_Y=1$ and \eqref{e:gX} reduces to $g_{X,\mbox{st}}=1+k_\alpha*{\tilde k_\alpha}/\rho_Y$. Thus $g_X\ge1$ and $g_X\not=1$ which is usually interpreted as $X$ being a model for clustering. This is of course also the situation if $g_Y\ge1$. However, such models for clustering may cause a large amount of overlap between the clusters unless $\rho_Y$ is small as compared to the band width $\alpha$. 

\subsection{The special setting of DSNCPs}\label{s:specialset}

When $Y$ is a DPP and we let $R_\beta(y)=|r_\beta(y,0)|^2$, we have 
 \begin{equation}\label{e:pcfDPP2} 
 g_{Y,\mbox{st}}(y)=1-R_\beta(y),\quad y\in\mathbb R^d,
 \end{equation} 
cf.\ \cite{LMR15}. Thus $g_Y\le1$, which reflects that a DPP is repulsive. From \eqref{e:gX} and \eqref{e:pcfDPP2} we get
\begin{equation}\label{e:gXDPP}
g_{X,\mbox{st}}(x)=1- k_\alpha*{\tilde k_\alpha}*R_\beta(x)+k_\alpha*{\tilde k_\alpha}(x)/\rho_Y,\quad x\in\mathbb R^d.
\end{equation}
This is in accordance to intuition: As $R_\beta$ increases, meaning  that $g_Y$ decreases and hence that $Y$ becomes more repulsive, it follows from \eqref{e:gXDPP} that $g_X$ decreases; and as the band width $\alpha$ tends to 0, we see that $g_{X,\mbox{st}}(x)$ tends to $g_{Y,\mbox{st}}(x)$ for every $x\in\mathbb R^d$. 
Below, we let $k_\alpha$ be the pdf of $N_d(0,\alpha^2I)$ and consider the pcfs and $K$-functions in the special cases of Gaussian/Ginibre-DPP-Thomas processes. 

Let $X$ be a Gaussian-DPP-Thomas process. Then $Y$ is a Gaussian DPP and
\begin{equation}\label{star}
R_\beta(y)=\exp\left(-2\|y/\beta\|^2\right),\quad y\in\mathbb R^d.
\end{equation}
Thus we obtain from \eqref{e:gXDPP} that $g_{X,\mbox{st}}(x)=g_{X,\mbox{iso}}(\|x\|)$ is isotropic with
\begin{equation}\label{e:pcf-Gaussian-DPP-Thomas}
g_{X,\mbox{iso}}(r)=1+\frac{\exp\left(-\frac{r^2}{4\alpha^2}\right)}{\left(4\pi\alpha^2\right)^{d/2}\rho_Y}-
\frac{\left(\beta^2/2\right)^{d/2}\exp\left(-\frac{r^2}{4\alpha^2+\beta^2/2}\right)}{\left(4\alpha^2+\beta^2/2\right)^{d/2}},\quad r>0.
\end{equation}
We have
\begin{equation}\label{star-star}
g_{X,\mbox{iso}}(r)\gtreqqless 1\ \Leftrightarrow\ r^2\lesseqqgtr
\frac{\ln\left(\rho_Y\left(\frac{2\pi\alpha^2\beta^2}{4\alpha^2+\beta^2/2}\right)^{d/2}\right)}{\frac{1}{4\alpha^2+\beta^2/2}-\frac{1}{4\alpha^2}},
\end{equation} 
where 
\[\frac{\ln\left(\rho_Y\left(\frac{2\pi\alpha^2\beta^2}{4\alpha^2+\beta^2/2}\right)^{d/2}\right)}{\frac{1}{4\alpha^2+\beta^2/2}-\frac{1}{4\alpha^2}}>0\]
since $\rho_Y\le\left(\pi\beta^2\right)^{-d/2}$. 
Furthermore, if $\omega_d$ denotes the volume of the $d$-dimensional unit ball and $F_{d/2}$ is the CDF of a gamma distribution with shape parameter $d/2$ and scale parameter $1$, we obtain 
\begin{equation}
K(r)=\omega_dr^d+\frac{1}{\rho_Y}F_{d/2}\left(\frac{r^2}{4\alpha^2}\right)
-\left(\pi\beta^2/2\right)^{d/2}F_{d/2}\left(\frac{r^2}{4\alpha^2+\beta^2/2}\right),\quad r>0.
\label{star-star-star}
\end{equation}

Let $X$ be a Ginibre-DPP-Thomas process. Then $Y$ is a scaled Ginibre point process which has some similarity to the Gaussian-DPP, since $\beta$ has the same range in the two processes and
\begin{equation}\label{e:star4}
R_\beta(y)=\exp(-|y/\beta|^2),\quad y\in\mathbb C,
\end{equation}
in the case of a scaled Ginibre point process. It thus follows from \eqref{e:pcfDPP2}, \eqref{star}, and \eqref{e:star4} that the pcfs of the scaled Ginibre point process and the Gaussian-DPP are of the same form, but $\beta^2$ in the scaled Ginibre point process corresponds to $\beta^2/2$ in the Gaussian-DPP. This shows that the scaled Ginibre point process is more repulsive than the Gaussian-DPP when using the same parameters $\rho_Y$ and $\beta$, and therefore it will be possible to obtain a larger repulsion between the clusters in a Ginibre-DPP-Thomas process than in a Gaussian-DPP-Thomas process. In fact, if $\beta=1/\sqrt{\pi\rho_Y}$ when $Y$ is a scaled Ginibre point process, then $Y$ is a most repulsive DPP in the sense of \cite{LMR15}. 
Because $\beta^2$ in the scaled Ginibre point process corresponds to $\beta^2/2$ in the Gaussian-DPP, \eqref{e:pcf-Gaussian-DPP-Thomas}--\eqref{star-star-star} give that 
$g_{X,\mbox{st}}(x)=g_{X,\mbox{iso}}(\|x\|)$ is isotropic with
\begin{equation*}\label{e:pcf-Ginibre-Thomas}
g_{X,\mbox{iso}}(r)=1+\frac{\exp\left(-r^2/(4\alpha^2)\right)}{4\pi\alpha^2\rho_Y}-
\frac{\beta^2\exp\left(-r^2/\left(4\alpha^2+\beta^2\right)\right)}{4\alpha^2+\beta^2},\quad r>0,
\end{equation*}
\[g_{X,\mbox{iso}}(r)\gtreqqless 1\ \Leftrightarrow\ r^2\lesseqqgtr
\frac{\ln\left(\rho_Y\frac{4\pi\alpha^2\beta^2}{4\alpha^2+\beta^2}\right)}{\frac{1}{4\alpha^2+\beta^2}-\frac{1}{4\alpha^2}},\]
and 
\[
K(r)=\pi r^2+\frac{1}{\rho_Y}\left(1-\exp\left(-\frac{r^2}{4\alpha^2}\right)\right)
-\pi\beta^2\left(1-\exp\left(-\frac{r^2}{4\alpha+\beta^2}\right)\right), \quad r>0.
\]

\begin{figure}[htp!]
\centering
\includegraphics[scale=1]{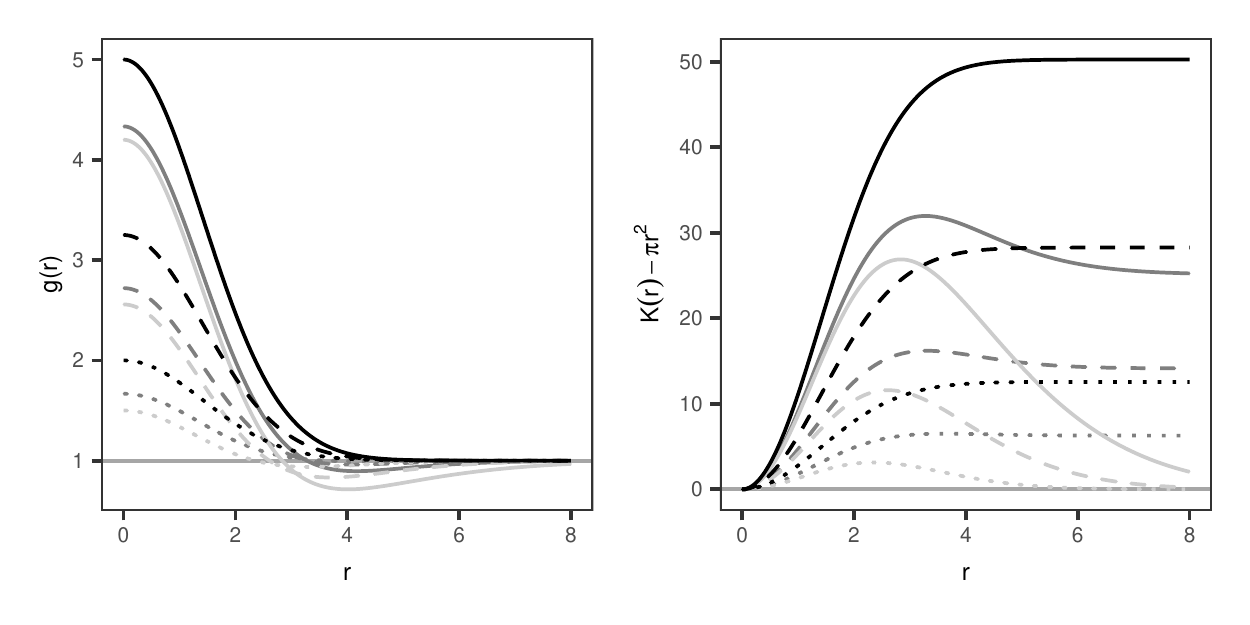}
\caption{Plots of $g_{X,\mbox{iso}}(r)$ (left) and $K(r)-\pi r^2$ (right) for the Thomas process (black curves), Gaussian-DPP-Thomas process (dark grey curves), and Ginibre-DPP-Thomas process (light grey curves) when $d=2$, $\alpha=1$, $\beta = 2,3,4$ (dotted, dashed, and solid curves, respectively), and $\rho_Y=1/(\beta^{2}\pi)$. The plots also shows the constant lines corresponding to the cases of a stationary Poisson process.}
\label{fig:pcf_K}
\end{figure}

Figure~\ref{fig:pcf_K} shows plots of $g_{X,\mbox{iso}}(r)$ and $K(r)-\pi r^2$ for Gaussian- and Ginibre-DPP-Thomas processes for different values of $\beta$ when $\rho_Y=1/(\pi\beta^{2})$ corresponds to the most repulsive case and without loss of generality we let $\alpha=1$.
For comparison the plots also include the case of a Thomas process with the same values for $\gamma$ and $\rho_Y$. Note that $K(r)-\pi r^2 = 0$ in case of a planar stationary Poisson process, and the figure shows that as $\beta$ increases, the processes behave less and less like a planar stationary Poisson process. The pair correlation functions in the cases of the DSNCP processes show an increasing degree of clustering at small scales and regularity at larger scales as $\beta$ increases, whereas Ripley's $K$-function only reveals an increasing degree of clustering. We furthermore see that the Ginibre-DPP-Thomas processes are overall more regular than the corresponding Gaussian-DPP-Thomas processes, especially at larger scales, which again reflects that the cluster centres are more regular in the Ginibre-DPP-Thomas processes. The considered Thomas processes are more clustered than the corresponding DSNCP processes and show no signs of repulsive behaviour.

\section{Statistical inference}\label{s:appl}

Suppose $d=2$, $W\subset\mathbb R^2$ is a bounded observation window, $X\cap W=\{x_1,\ldots,x_n\}$ is a point pattern data set, and we want to fit a parametric DSNCP model given by either the Gaussian-DPP-Thomas or Ginibre-DPP-Thomas process. There is a trade-off between $\rho_Y$ and $\gamma$ because of the relation $\rho_X=\gamma\rho_Y$. 
Therefore, when modelling the data, we choose to let $\rho_Y=1/(\pi\beta^2)$, which means that $Y$ will be as repulsive as possible. That is, $\gamma>0$ and $\theta=(\alpha,\beta)\in(0,\infty)^2$ are the unknown parameters.

\subsection{Estimation}
\label{sec:estimation}

Likelihood based inference is complicated because of the unobserved process of cluster centres.\\ \cite{MW2004} showed how a missing data MCMC approach can be used for maximum likelihood estimation in the special case of the Thomas process, and it may be simpler but still rather complicated to use a MCMC Bayesian setting along similar lines as in \cite{Beraha22}. 
We propose instead to exploit the parametric expressions of the intensity  and of the pcf or $K$-function given  in Section~\ref{s:specialset} when estimating $\gamma$ and $\theta$. In this paper, we use a minimum contrast procedure and leave it for future research to investigate the alternative approaches of composite likelihood \citep{Guan} and Palm likelihood \citep{TanakaEtAl} using the expressions of $\rho_X$ and $g_{X,\mbox{st}}$, see the review in \cite{JMRW} and the references therein.

Specifically, we use a minimum contrast procedure for estimating $\theta$, where it is preferable to consider Ripley's $K$-function, since it is easier to estimate $K$ than $g_{X,\mbox{st}}$ by non-parametric methods, see e.g.\ \cite{MW2004}. Since $K$ does not depend  on $\gamma$, we need to estimate $\gamma$
 separately. 
 Writing $K=K_\theta$ to stress the dependence of $\theta$ and $\hat K$ for a non-parametric estimate based on $\{x_1,\ldots,x_n\}$, the minimum contrast estimate of $\theta$ is given by 
\begin{equation*}
\hat{\theta}=\underset{\theta}{\operatorname{arg\,min}}\left \{\int_{r_{\mathrm{min}}}^{r_{\mathrm{max}}} |\hat{K}(r)^q - K_\theta(r)^q|^p\,\text{d}r\right \}
\end{equation*}
where we use the \texttt{R}-package \texttt{spatstat} \citep{BRT2015} for calculating $\hat K$ and the minimum contrast estimate by using default settings for
the choice of $r_{\mathrm{min}}, r_{\mathrm{max}}, q,$ and $p$. Finally, having estimated $\theta$, we estimate $\gamma$ from the unbiased estimation equation $\rho_X=\gamma/(\pi\beta^2)=\gamma\rho_Y=n/|W|$.

\subsection{Model checking}
\label{sec:model_checking}

When checking a fitted model, we prefer to use other functional summary statistics than $\hat K$ since this was used as part of the estimation procedure. The standard is to consider empirical estimates of theoretical functions known as the empty space function (or spherical contact function) $F$, the nearest-neighbour function $G$, and the $J$-function, which are defined for a stationary point process $X$ as follows. Consider any number $r>0$ and an arbitrary point $u\in\mathbb R^d$. Then,
\begin{align*}
F(r)&= \mathrm P({\mathrm{dist}}(X,u)\le r),\\
G(r)&= \mathrm P({\mathrm{dist}}(X\setminus\{u\},u)\le r\,|\,u\in X),\\
J(r)&= (1-G(r))/(1-F(r)).
\end{align*}
Here $J$ is only defined for $F(r)<1$, ${\mathrm{dist}}(X,u)=\inf\{r>0\,|\, b(u,r)\cap X\not=\emptyset\}$ is the distance from $u$ to 
$X$, and in the definition of $G$ when conditioning on $u\in X$ it means that $X\setminus\{u\}$ follows the reduced Palm distribution of $X$ at $u$, see e.g.\ \citet{MW2004}. Since $X$ is stationary, the definitions of $F, G,$ and $J$ do not depend on the choice of $u$. 

\begin{figure}[htp!]
\centering
\includegraphics[scale=1]{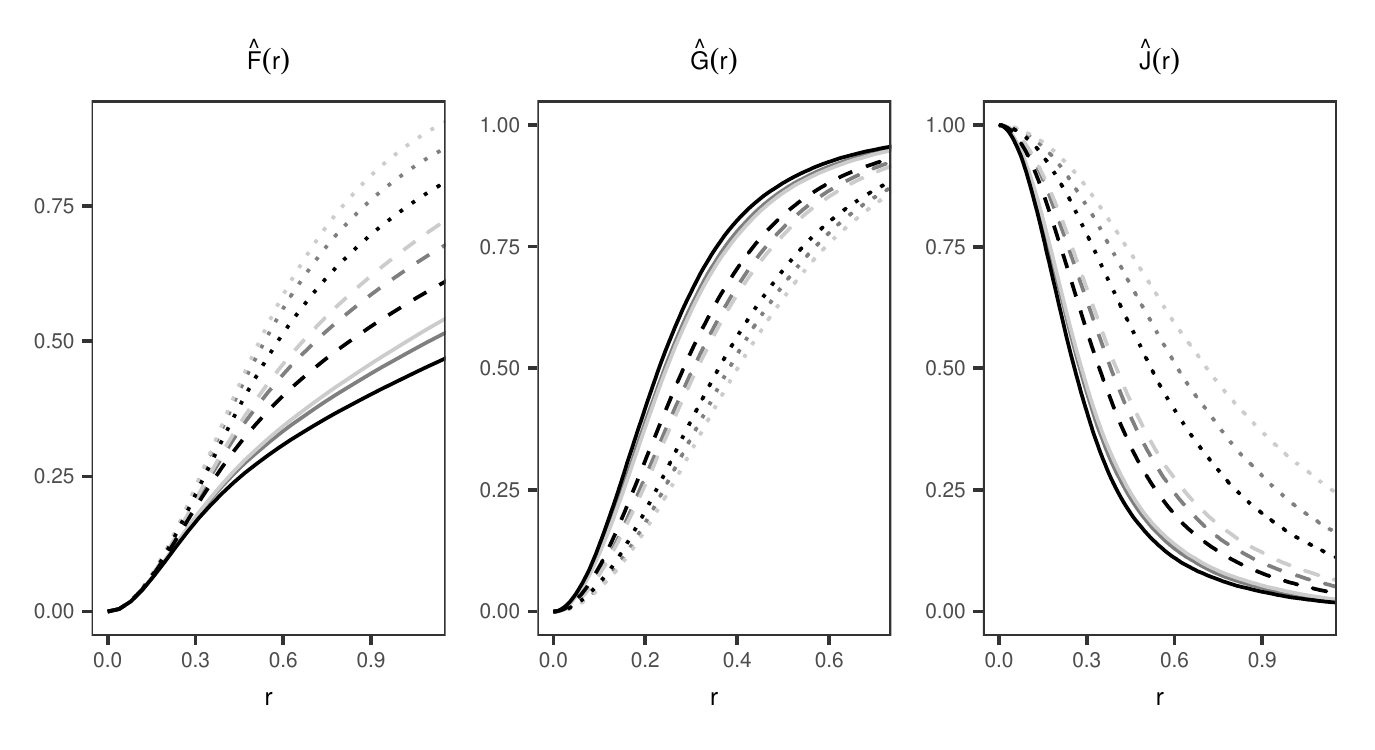}
\caption{Means of $\hat{F}, \hat{G}$, and $\hat{J}$ calculated from 500 simulations of Thomas processes (black curves), Gaussian-DPP-Thomas processes (dark grey curves), and Ginibre-DPP-Thomas processes (light grey curves) on a square with side lengths 20. In all types of processes, $\alpha = 1$, $\beta = 2,3,4$ (dotted, dashed, and solid curves, respectively), $\rho_Y=1/(\pi\beta^2)$, and $\gamma = 1/\rho_Y$. Note that $\beta$ is not a parameter of the Thomas process and is thus only used to calculate $\rho_{Y}$ in this case.}
\label{fig:F_G_J}
\end{figure}

We have not been able to derive the expressions of $F$, $G$, and $J$ for Gaussian- and Ginibre-DPP-Thomas processes; to the best of our knowledge, these expressions are not even known for a Thomas process. We refer to empirical estimates of these theoretical functions as functional summary statistics and use the relevant functions in \texttt{spatstat} to calculate such non-parametric estimates (always using the default settings including settings which account for boundary effects). Figure~\ref{fig:F_G_J} concerns means of non-parametric estimates $\hat F$, $\hat G$, and $\hat J$ calculated from simulations of Thomas processes, Ginibre-DPP-Thomas processes and Gaussian-DPP-Thomas processes for the parameters stated in the caption.
In agreement with Figure~\ref{fig:pcf_K} the plots show an increasing degree of clustering as $\beta$ increases and that the Thomas processes are more clustered than the corresponding DSNCP processes. The plots also indicate that the Gaussian-DPP-Thomas processes are more clustered and exhibit more empty space than the corresponding Ginibre-DPP-Thomas processes, and the difference becomes more apparent as $\beta$ decreases. 

In order to validate a fitted model, we use 95\% global envelopes and global envelope tests based on the extreme rank length as described in \citet{GET2017}, \citet{GET2018}, and \citet{GETinR}, which is implemented in the \texttt{R}-package \texttt{GET} \citep{GETinR}. These envelopes are based on functional summary statistics calculated from a number of simulations under the fitted model. We use $2499$ simulations as recommended in the above references. 

\subsection{An application example}\label{sec:dataex}

\begin{figure}[ht!]
\centering
\includegraphics[scale=1]{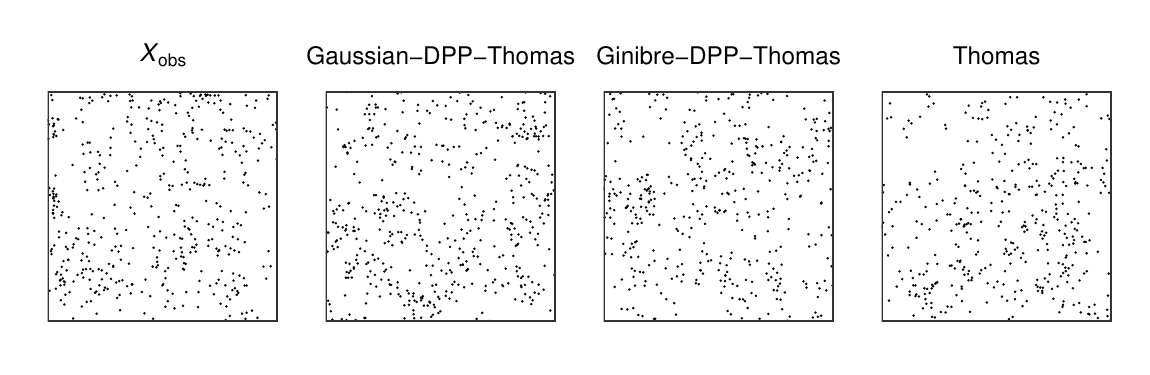}
\caption{Plots of the whiteoak point pattern ($x_{\mathrm{obs}}$) and a simulation from fitted models of the type stated at the top of each plot.}
\label{fig:dataexample_pp}
\end{figure}

\begin{figure}[htp!]
\centering
\includegraphics[scale=1]{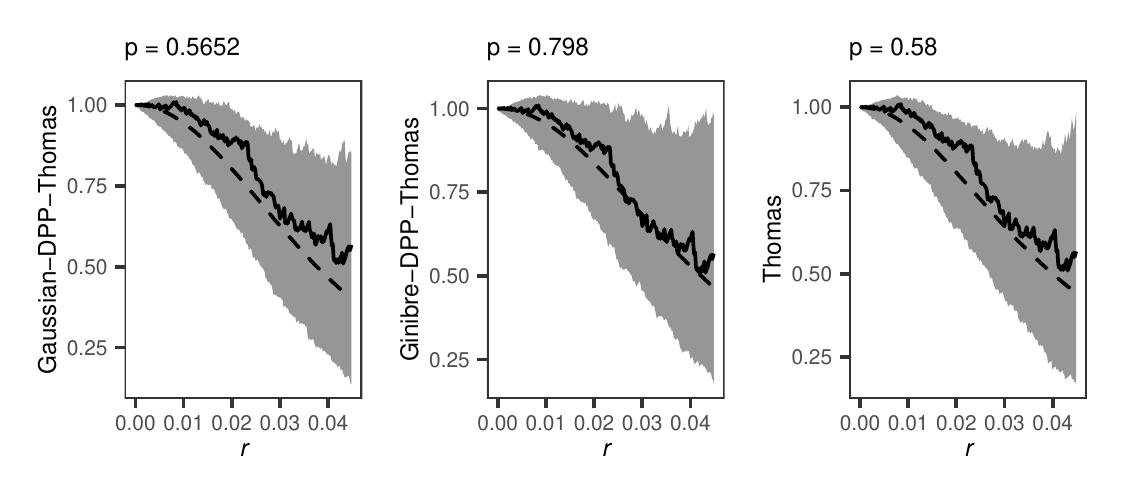}
\caption{Plots of 95\% global envelopes (grey area) and tests ($p$-value stated above each plot) based on $\hat{J}(r)$ from 2499 simulations from the fitted model stated to the left of each plot. The solid curves show $\hat{J}$ for $X_{obs}$, and the dashed curves show the mean of $\hat{J}$ calculated from the simulations.}
\label{fig:dataexample_GET}
\end{figure}

The first point pattern in Figure~\ref{fig:dataexample_pp} shows the positions of 448 white oak trees in a square region (scaled to a unit square) of Lansing Woods, Clinton County, Michigan USA, which is part of the \texttt{lansing} data set which is available in \texttt{spatstat}.
We will refer to this point pattern as $x_{\mathrm{obs}}$. By using the method of minimum contrast estimation as described in Section~\ref{sec:estimation}, we fitted a Gaussian-DPP-Thomas process, Ginibre-DPP-Thomas process, and Thomas process to $x_{\mathrm{obs}}$. The obtained estimates are given in Table~\ref{tab:parameters}.
\begin{table}[ht!]
\caption{Estimated parameters when fitting models to $x_{\mathrm{obs}}$. Note that the Thomas process does not have the parameter $\beta$.}
\centering
\begin{tabular}{lcccc} 
Model & $\beta$ & $\rho_Y$ & $\gamma$ & $\alpha$\\
\hline
Gaussian-DPP-Thomas & 0.05 & 105.36 & 4.25 & 0.03\\
Ginibre-DPP-Thomas & 0.09 & 35.32 & 12.68 & 0.05\\
Thomas & - & 204.11 & 2.19 & 0.03\\
\end{tabular}
\label{tab:parameters}
\end{table}
We see that the fitted DPP-Thomas processes expect much fewer clusters than the Thomas process and thus also more points in each cluster. As we expected, the fitted Ginibre-DPP-Thomas process is the one which expects the fewest clusters. Because of its expected 35 clusters with about 12 points on average in each it also seems to be a more sensible cluster process model than the other processes, which have many clusters with only a few points in each cluster. Figure~\ref{fig:dataexample_pp} also shows a realization of each fitted model. The behaviour of these realizations are apparently in good agreement with $x_{\mathrm{obs}}$. In order to check whether the models fit to data, we made 95\% global envelope tests as described in Section~\ref{sec:model_checking}. Figure~\ref{fig:dataexample_GET} shows the results, which indicate that all three models fit well, but the Ginibre-DPP-Thomas process has a much higher $p$-value than the other two processes. 

In connection with this paper we considered over 100 examples of point pattern data sets and found 20 point patterns which the considered DSNCP models describe well, including the application example in this section. For all of these we also found that the Thomas process fits well, but that the fitted Thomas process models expected more clusters than the corresponding fitted DSNCP models. Thus the situation exemplified in this section where all three of the considered models can be used to model data but the DSNCP models expect fewer clusters appears to be a typical situation.

\section{Simulation study}\label{sec:GETsimstudy}
Section~\ref{sec:dataex} suggests that it may be difficult to distinguish between realizations of Thomas, Gaussian-DPP-Thomas, and Ginibre-DPP-Thomas processes. To investigate this further, we in this section describe a simulation study where we considered the parameters $\rho_Y=10,30,50$, $\gamma = 10,30,50$, and $\alpha = 0.03, 0.04, 0.05$ for the three considered cluster point process models (in the DPP-Thomas processes we as always used the relation $\rho_Y=1/(\pi\beta^2)$ or equivalently $\beta = \sqrt{1/(\pi\rho_{Y})}$). The values of $\alpha$ are like those in Table~\ref{tab:parameters}, and the values of $\rho_Y$ and $\gamma$ are like those from the fitted Ginibre-DPP-Thomas process in Table~\ref{tab:parameters}. For each combination of parameters and each model we made 100 simulations on a unit square. For each of these simulations, we fitted the two models which were not the true one and made a global envelope test as described in Section~\ref{sec:dataex} for validating the fitted models. Since this simulation study is time consuming, we only used 1999 simulations to calculate each global envelope test in order to save some time, but this is still in agreement with the recommendations regarding the number of simulations in global envelope tests.

\begin{figure}[ht!]
\centering
\includegraphics[scale=1]{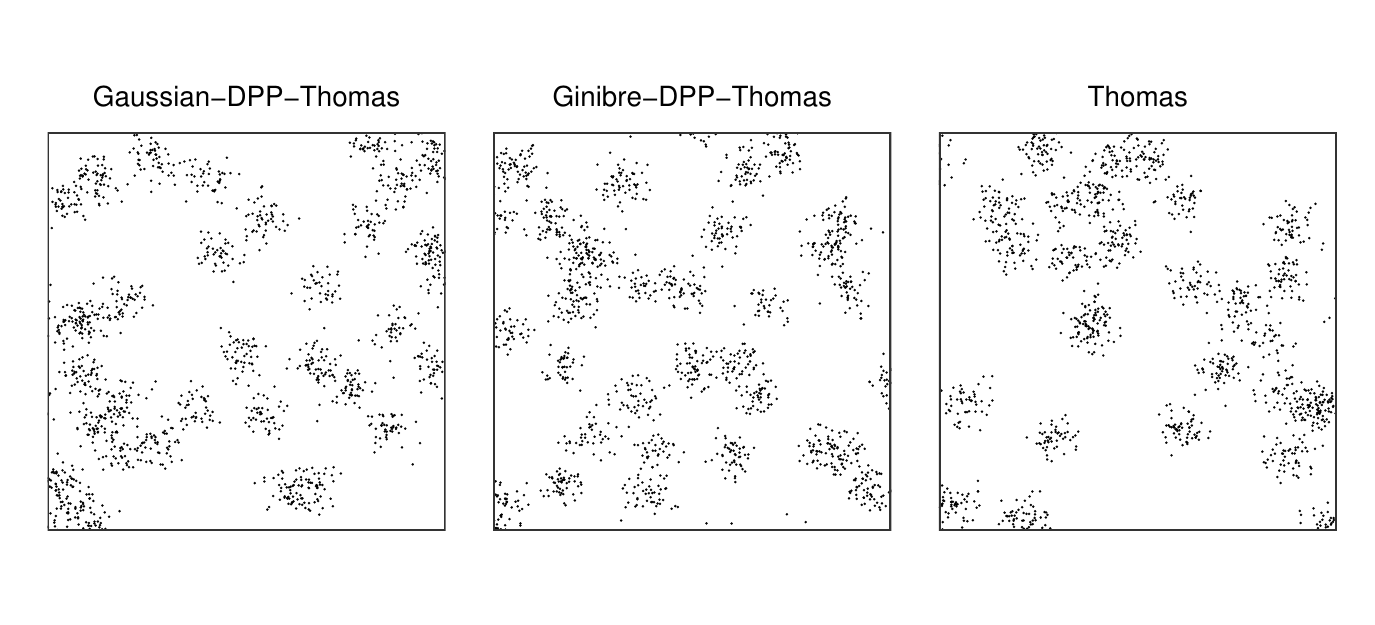}
\caption{Simulation of a realization under the model stated at the top with parameters $\alpha = 0.03$, $\gamma = 50$, and $\rho_{Y}=1/(\pi\beta^2)=30$.}
\label{fig:sims_GETsimstudy}
\end{figure}

Table~\ref{tab:rejectrate} shows the proportion of tests which yielded a $p$-value below $0.05$ for each combination of parameters, true model, and fitted model.
\begin{table}[ht!]
\caption{Table of the proportion of global envelope tests in the simulation study for for which the $p$-value was below $0.05$. Concerning the column with the fitted model, for short the models Thomas process, Gaussian-DPP-Thomas process, and Ginibre-DPP-Thomas process are written as Thomas, Gaussian, and Ginibre, respectively.}
\centering
\begin{tabular}{|l|l|lll|lll|lll|}
  \hline
  && \multicolumn{3}{c|}{$\rho_{Y}=10$} & \multicolumn{3}{c|}{$\rho_{Y}=30$} & \multicolumn{3}{c|}{$\rho_{Y}=50$}\\ \hline
  Fitted model: & & $\gamma=10$ & $\gamma=30$ & $\gamma=50$ & $\gamma=10$ & $\gamma=30$ & $\gamma=50$ & $\gamma=10$ & $\gamma=30$ & $\gamma=50$ \\ 
  \hline
  \multicolumn{11}{|c|}{True model is a Ginibre-DPP-Thomas process}\\
  \hline
 & $\alpha=0.03$ & 0.02 & 0.35 & 0.37 & 0.05 & 0.56 & 0.79 & 0.06 & 0.26 & 0.52 \\ 
Thomas & $\alpha=0.04$ & 0.03 & 0.11 & 0.30 & 0.02 & 0.06 & 0.22 & 0.05 & 0.06 & 0.00 \\ 
& $\alpha=0.05$ & 0.01 & 0.09 & 0.20 & 0.02 & 0.01 & 0.01 & 0.03 & 0.01 & 0.03 \\ 
   \hline
   &$\alpha=0.03$ & 0.02 & 0.09 & 0.14 & 0.02 & 0.33 & 0.55 & 0.05 & 0.33 & 0.51 \\ 
  Gaussian& $\alpha=0.04$ & 0.06 & 0.09 & 0.10 & 0.02 & 0.11 & 0.24 & 0.04 & 0.10 & 0.06 \\ 
  &$\alpha=0.05$ & 0.02 & 0.05 & 0.18 & 0.05 & 0.03 & 0.05 & 0.03 & 0.02 & 0.02 \\ 
   \hline
   \multicolumn{11}{|c|}{True model is a Gaussian-DPP-Thomas process}\\
  \hline
  &$\alpha=0.03$ & 0.02 & 0.17 & 0.22 & 0.03 & 0.20 & 0.19 & 0.07 & 0.09 & 0.13 \\ 
  Thomas &$\alpha=0.04$ & 0.02 & 0.06 & 0.09 & 0.04 & 0.02 & 0.05 & 0.02 & 0.04 & 0.02 \\ 
  &$\alpha=0.05$ & 0.02 & 0.06 & 0.10 & 0.03 & 0.01 & 0.00 & 0.01 & 0.03 & 0.04 \\ 
   \hline
   &$\alpha=0.03$ & 0.02 & 0.10 & 0.08 & 0.18 & 0.31 & 0.44 & 0.21 & 0.35 & 0.44 \\ 
  Ginibre&$\alpha=0.04$ & 0.03 & 0.07 & 0.10 & 0.09 & 0.15 & 0.15 & 0.04 & 0.14 & 0.15 \\ 
  &$\alpha=0.05$ & 0.04 & 0.06 & 0.06 & 0.05 & 0.03 & 0.02 & 0.04 & 0.02 & 0.07 \\ 
   \hline
   \multicolumn{11}{|c|}{True model is a Thomas process}\\
  \hline
  &$\alpha=0.03$ & 0.08 & 0.19 & 0.25 & 0.25 & 0.45 & 0.60 & 0.19 & 0.48 & 0.56 \\ 
  Ginibre &$\alpha=0.04$ & 0.06 & 0.11 & 0.12 & 0.07 & 0.15 & 0.18 & 0.06 & 0.09 & 0.08 \\ 
  &$\alpha=0.05$ & 0.03 & 0.05 & 0.05 & 0.03 & 0.06 & 0.10 & 0.02 & 0.05 & 0.02 \\ 
   \hline
   &$\alpha=0.03$ & 0.02 & 0.08 & 0.06 & 0.06 & 0.07 & 0.15 & 0.08 & 0.08 & 0.19 \\ 
  Gaussian & $\alpha=0.04$ & 0.03 & 0.07 & 0.04 & 0.06 & 0.06 & 0.06 & 0.04 & 0.06 & 0.04 \\ 
  & $\alpha=0.05$ & 0.02 & 0.04 & 0.01 & 0.05 & 0.02 & 0.03 & 0.02 & 0.04 & 0.05 \\ 
   \hline
\end{tabular}
\label{tab:rejectrate}
\end{table}
We overall see that in order to distinguish between the models, the ideal situation is when $\gamma$ is large and $\alpha$ is small, meaning that the realization consists of small clusters with many points in each. Overall, it also seems to be an advantage if there is a moderate number of clusters since the rejection rates are generally higher when $\rho_{Y}=30$, especially for small $\alpha$. It appears to be most difficult to distinguish between the Gaussian-DPP-Thomas process and the two remaining processes, especially the Thomas process, whereas it is easier to distinguish between Thomas and Ginibre-DPP-Thomas processes. Figure~\ref{fig:sims_GETsimstudy} shows a realization under each model with parameters $\alpha = 0.03$, $\gamma = 50$, and $\rho_{Y}=30$, which the simulation study suggests is a good situation when it comes to distinguishing between the models.

We also used this simulation study to investigate the apparent tendency for fitted Thomas processes to expect more clusters than fitted DPP-Thomas processes. Table~\ref{tab:mean_rhoY} shows the mean of the fitted value of $\rho_{Y}$ divided by the value of $\rho_Y$ in the true model for each combination of parameters, model, and fitted model.
\begin{table}[ht!]
\caption{Table of the mean of the fitted value of $\rho_Y$ divided by the value of $\rho_Y$ in the true model. Concerning the column with the fitted model, for short the models Thomas process, Gaussian-DPP-Thomas process, and Ginibre-DPP-Thomas process are written as Thomas, Gaussian, and Ginibre, respectively.}
\centering
\begin{tabular}{|l|l|lll|lll|lll|}
  \hline
  && \multicolumn{3}{c|}{$\rho_{Y}=10$} & \multicolumn{3}{c|}{$\rho_{Y}=30$} & \multicolumn{3}{c|}{$\rho_{Y}=50$}\\ \hline
  Fitted model: & & $\gamma=10$ & $\gamma=30$ & $\gamma=50$ & $\gamma=10$ & $\gamma=30$ & $\gamma=50$ & $\gamma=10$ & $\gamma=30$ & $\gamma=50$ \\ 
  \hline
  \multicolumn{11}{|c|}{True model is a Ginibre-DPP-Thomas process}\\
  \hline
&$\alpha = 0.03$ & 1.92 & 1.85 & 1.84 & 2.83 & 2.77 & 2.78 & 3.71 & 3.55 & 3.54 \\ 
 Thomas & $\alpha =0.04$ & 2.10 & 2.10 & 2.08 & 3.78 & 3.52 & 3.65 & 5.11 & 4.71 & 4.85 \\ 
  &$\alpha =0.05$ & 2.50 & 2.23 & 2.54 & 4.93 & 4.66 & 4.32 & 6.73 & 6.50 & 6.68 \\ 
   \hline
   &$\alpha =0.03$ & 1.26 & 1.22 & 1.22 & 1.60 & 1.57 & 1.57 & 1.98 & 1.91 & 1.90 \\ 
  Gaussian&$\alpha =0.04$ & 1.32 & 1.32 & 1.31 & 2.03 & 1.89 & 1.96 & 2.63 & 2.41 & 2.50 \\ 
  &$\alpha =0.05$ & 1.49 & 1.35 & 1.51 & 2.55 & 2.42 & 2.25 & 3.41 & 3.29 & 3.38 \\ 
   \hline
   \multicolumn{11}{|c|}{True model is a Gaussian-DPP-Thomas process}\\
  \hline
  &$\alpha =0.03$ & 1.52 & 1.59 & 1.71 & 1.89 & 1.93 & 1.81 & 2.03 & 2.02 & 2.04 \\ 
  Thomas&$\alpha =0.04$ & 1.72 & 1.73 & 1.61 & 2.00 & 2.01 & 2.12 & 2.41 & 2.19 & 2.17 \\ 
  &$\alpha =0.05$ & 1.87 & 1.95 & 1.98 & 2.41 & 2.36 & 2.18 & 2.31 & 2.49 & 2.45 \\ 
   \hline
&$\alpha =0.03$ & 0.85 & 0.89 & 0.93 & 0.70 & 0.71 & 0.67 & 0.59 & 0.59 & 0.59 \\ 
  Ginibre &$\alpha =0.04$ & 0.85 & 0.86 & 0.81 & 0.61 & 0.61 & 0.63 & 0.55 & 0.51 & 0.52 \\ 
  &$\alpha =0.05$ & 0.83 & 0.86 & 0.87 & 0.62 & 0.60 & 0.57 & 0.45 & 0.49 & 0.49 \\ 
   \hline
   \multicolumn{11}{|c|}{True model is a Thomas process}\\
  \hline
  &$\alpha =0.03$ & 0.70 & 0.65 & 0.64 & 0.44 & 0.45 & 0.43 & 0.37 & 0.38 & 0.38 \\ 
  Ginibre &$\alpha =0.04$ & 0.66 & 0.59 & 0.64 & 0.44 & 0.42 & 0.43 & 0.34 & 0.34 & 0.35 \\ 
  &$\alpha =0.05$ & 0.59 & 0.59 & 0.61 & 0.38 & 0.40 & 0.40 & 0.33 & 0.31 & 0.33 \\ 
   \hline
   &$\alpha =0.03$ & 0.83 & 0.77 & 0.75 & 0.65 & 0.67 & 0.62 & 0.62 & 0.64 & 0.64 \\ 
  Gaussian&$\alpha =0.04$ & 0.81 & 0.72 & 0.78 & 0.71 & 0.66 & 0.69 & 0.64 & 0.65 & 0.66 \\ 
  &$\alpha =0.05$ & 0.76 & 0.76 & 0.78 & 0.67 & 0.71 & 0.70 & 0.69 & 0.64 & 0.68 \\ 
   \hline
\end{tabular}
\label{tab:mean_rhoY}
\end{table}
We see that when the true model is a DPP-Thomas process, the fitted Thomas processes expect more clusters than the true model, especially when the true model is a Ginibre-DPP-Thomas process; this behaviour gets more extreme as $\alpha$ and $\rho_Y$ increases. This is also the behaviour of the fitted Gaussian-DPP-Thomas processes when the true model is a Ginibre-DPP-Thomas process, although it is not as extreme as for the fitted Thomas processes. If the true model is a Thomas process, we similarly see that the fitted DPP-Thomas processes expect fewer clusters than the true model, especially the Ginibre-DPP-Thomas process; this behaviour gets more extreme as $\rho_Y$ increases, whereas in this case it seems that $\alpha$ has only little influence on this behaviour. This is also the behaviour of the fitted Ginibre-DPP-Thomas processes when the true model is a Gaussian-DPP-Thomas process. The parameter $\gamma$ in the true model has no apparent effect on the expected number of clusters in the fitted models. 

\section{Conclusion}\label{sec:conclusion}
We have presented the new class of cluster point process models called determinantal shot noise Cox processes which have repulsion between cluster centres. For the two special cases which we have called Gaussian-DPP-Thomas processes and Ginibre-DPP-Thomas processes we have derived closed form expressions for the pair correlation function and Ripley's $K$-function. The ability to actually derive such closed form parametric expressions for these theoretical summary functions is a huge advantage compared to using Markov point processes for the cluster centres, which has previously been done for cluster point processes with repulsion between clusters, since easy and fast parameter estimation can then be achieved with the method of minimum contrast estimation or other methods based on the pair correlation or $K$-function, cf.\ Section~\ref{sec:estimation}. 

We have seen that the fitted DPP-Thomas process models in Sections~\ref{sec:dataex} and \ref{sec:GETsimstudy} expect much fewer clusters than a Thomas process and thus they also expect much more points in each cluster, especially the Ginibre-DPP-Thomas model. In many situations it will be intuitively more pleasing to fit a cluster point process with few clusters consisting of many points compared to many clusters consisting of very few points. We have also seen through a simulation study that the ideal situation for distinguishing between the considered three types of cluster point process models is if the realization has small clusters with many points in each.

\section*{Acknowledgements}
We would like to thank Rasmus Plenge Waagepetersen for providing us with many examples of point pattern data sets which we considered in connection with the application example.

\section*{Appendix A: Definition of a DPP and some properties}

Let $Y$ be a simple point process defined on $\mathbb R^d$ and $c$ be a complex function defined on $\mathbb R^d\times\mathbb R^d$ so that 
for every integer $n>0$ and pairwise disjoint bounded Borel sets $A_1,\ldots,A_n\subset\mathbb R^d$, we have
\[
\mathrm E\left[Y(A_1)\cdots Y(A_n)\right]=\int_{A_1}\cdots\int_{A_n}\mbox{det}\{c(u_i,u_j)\}_{i,j=1,\ldots,n}\,\mathrm du_1\cdots\,\mathrm du_n<\infty
\]
where $\mbox{det}\{c(u_i,u_j)\}_{i,j=1,\ldots,n}$ is the determinant of the $n\times n$ matrix with $ij$'th entry $c(u_i,u_j)$. Then
\cite{Macchi:75} defined $Y$ to be a DPP with kernel $c$. Note that 
 $Y$ must be locally finite and the function 
\begin{equation}\label{e:def-DPP}
\rho^{(n)}(u_1,\ldots,u_n)=\mbox{det}\{c(u_i,u_j)\}_{i,j=1,\ldots,n}
\end{equation}
 is the so-called $n$'th
order intensity function $\rho^{(n)}$ of $Y$. 

In fact for the DPP $Y$, its distribution is unique and completely characterized by the intensity functions of all order, cf.\ Lemma~4.2.6 in \cite{Hough:etal:09}. Thus stationarity of $Y$ is equivalent to that $\rho^{(n)}(u_1,\ldots,u_n)=\rho^{(n)}(u_1+v,\ldots,u_n+v)$ for all $v\in\mathbb R^d$ and (Lebesgue almost) all $u_1,\ldots,u_n\in\mathbb R^d$, and isotropy of $Y$ means that $\rho^{(n)}(u_1,\ldots,u_n)=\rho^{(n)}(\mathcal O u_1,\ldots,\mathcal O u_n)$ for all $n\times n$ rotations matrices $\mathcal O$ and (Lebesgue almost) all $u_1,\ldots,u_n\in\mathbb R^d$. 

For later use, consider any numbers $\beta>0$ and $0\le p\le 1$, and the scaled point process $\beta Y=\{\beta y\mid y\in Y\}$. 
Let $Y_{\beta,p}$ be an independent $p$-thinning of $\beta Y$  (that is, the points in $\beta Y$ are independently retained with probability $p$ and $Y_{\beta,p}$ consists of those retained points). It is easily seen that $Y_{\beta,p}$ is a DPP with kernel
\begin{equation}\label{e:ccccccccc}
c_{\beta,p}(u,v)=(p/\beta)^d c(u/\beta,v/\beta).
\end{equation}   

\section*{Appendix B: Some results for the scaled Ginibre point processes}

In the following assume $d=2$ and $Y$ is a standard Ginibre point process as defined in Section~\ref{sec:DSNCP}, so we identify $\mathbb R^2$ with the complex plane $\mathbb C$. 
Let $Y_{\beta,p}$ be as above and let $\lambda=\rho_{Y_{\beta,p}}$ be its intensity. By \eqref{e:ccccccccc}, $Y_{\beta,p}$ is the DPP 
with kernel
\[
c_{\beta,p}(u,v)=\lambda\exp((u\overline {v}-|u|^2/2-|v|^2/2)/\beta^2),\quad u,v\in\mathbb C,
\]
and $\lambda=(p/\beta)^2/\pi$.
In Section~\ref{sec:DSNCP} we used the variation dependent parametrization $(\rho_{Y_{\beta,p}},\beta)$, which is in one-to-one correspondence to 
$(\beta,p)$. For the following it is convenient to let $\nu=p^2$ and use the variation independent parametrization $(\nu,\lambda)\in(0,1]\times(0,\infty)$, which is also in one-to-one correspondence to 
$(\beta,p)$. Using this parametrization, with a slight abuse of notation we write $Y_{\nu,\lambda}$ for the DPP $Y_{\beta,p}$ and
\begin{equation}\label{e:cGDPP}
c_{\nu,\lambda}(u,v)=\lambda\exp\left((\lambda\pi/\nu)(u\overline {v}-|u|^2/2-|v|^2/2)\right)
\end{equation}
for its kernel. 
 
\subsection*{Appendix B.1: Invariance under isometries} 

Below we show that the $n$'th order intensity function is invariant under translations and rotations, and therefore $Y$ is stationary and isotropic. In the same way, it can be shown that $\rho^{(n)}$ is invariant under reflections and glide reflections. So the distribution of $Y$ is invariant under isometries (mappings of the form $z\to az+b$ and $z\to a\bar z +b$ where $a,b\in\mathbb C$ with $|a|=1$; these mappings correspond to translations, rotations, reflections, and glide reflections).

Denote $S_n$ the set of all permutations of $\{1,2,\ldots,n\}$ and $\text{sgn}(\sigma)$ the sign of a permutation $\sigma\in S_n$. From \eqref{e:def-DPP} and \eqref{e:cGDPP} we get 
\begin{align*}
\rho^{(n)}(u_1,\ldots,u_n)&=\sum_{\sigma\in S_n}\text{sgn}(\sigma)\prod_{i = 1}^{n}c_{\nu,\lambda}(u_{i},u_{\sigma(i)})\\
&=\lambda^{n}\sum_{\sigma\in S_{n}}\text{sgn}(\sigma)\exp\left ((\lambda\pi/\nu)\sum_{i=1}^{n}\left (u_{i}\overline{u}_{\sigma(i)}-|u_{i}|^{2}/2-|u_{\sigma(i)}|^{2}/2\right )\right )\\
&=\lambda^{n}\sum_{\sigma\in S_{n}}\text{sgn}(\sigma)\exp\left ((\lambda\pi/\nu)\sum_{i=1}^{n}\left (u_{i}\overline{u}_{\sigma(i)}-|u_{i}|^{2}\right )\right ).
\end{align*}
Hence, for any $a,b,u_1,\ldots,u_n$, a straightforward calculation gives
\[\rho^{(n)}(au_1+b,\ldots,au_n+b)=\rho^{(n)}(u_1,\ldots,u_n),\]
so $\rho^{(n)}$ is invariant under translations and rotations.

\subsection*{Appendix B.2: Spectral decompositions}

Spectral representations of the kernel restricted to compact regions are needed for simulation 
as well as other purposes, cf.\ \cite{LMR15}. The simplest case occurs when we consider $Y_{\nu,\lambda}$ restricted to a closed disc around zero. So for $r>0$, let $b(0,r)\subset \mathbb{C}$ be the closed disk around zero with radius $r\in(0,\infty)$ and $Y_{\nu,\lambda,r}=Y_{\nu,\lambda}\cap b(0,r)$ the restriction of $Y_{\nu,\lambda}$ to $b(0,r)$. Because $Y_{\nu,\lambda}$ is a DPP, $Y_{\nu,\lambda,r}$ is a DPP with kernel
\[c_{\nu,\lambda,r}(u,v)=\begin{cases}c_{\nu,\lambda}(u,v) &\mbox{if }(u,v)\in b(0,r),\\ 0 &\mbox{otherwise}.\end{cases}\]

The integral operator corresponding to the kernel $c_{\nu,\lambda}$ has only one eigenvalue, namely $\nu$, and the
eigenfunctions are
\begin{equation*}
\phi_{\nu,\lambda}^{i}(u)= \frac{\sqrt{\lambda}(\lambda\pi)^{(i-1)/2}}{\sqrt{(i-1)!\nu^{i}}}\exp(-\lambda\pi|u|^{2}/(2\nu))u^{i-1},\quad u\in\mathbb{C},\quad i=1,2,\ldots.
\end{equation*}
This follows easily by exploiting the moment properties of two independent zero-mean complex normally distributed random variables and the definition  of the complex exponential function ($\exp(z)=\sum_{k=0}^\infty z^k/k!$ for $z\in\mathbb C$) for the term $\exp((\lambda\pi/\nu)u\overline v)$ in \eqref{e:cGDPP}. In other words, we have
 the spectral representation 
\begin{equation*}
c_{\nu,\lambda}(u,v)=\sum_{i=1}^{\infty}\nu\phi_{\nu,\lambda}^{i}(u)\overline{\phi_{\nu,\lambda}^{i}(v)}.
\end{equation*} 
Similarly, we see that the integral operator corresponding to $c_{\nu,\lambda,r}$ has eigenfunctions
\begin{equation*}
\phi_{\nu,\lambda,r}^{i}(u) = 
\phi_{\nu,\lambda}^{i}(z)/\sqrt{F_i(\lambda\pi r^2/\nu)},\quad u\in b(0,r),\quad i=1,2,\ldots,
\end{equation*}
with corresponding eigenvalues 
\begin{equation}\label{eq:eigen}
\xi_{\nu,\lambda,r}^{i} = 
\nu F_i(\lambda\pi r^2/\nu), \quad i=1,2,\ldots, 
\end{equation}
and the spectral representation is
\begin{equation*}
c_{\nu,\lambda,r}(u,v)=\sum_{i=1}^{\infty}\xi_{\nu,\lambda,r}^{i}\phi_{\nu,\lambda,r}^{i}(u)\overline{\phi_{\nu,\lambda,r}^{i}(v)}.
\end{equation*} 

\section*{Appendix C: Simulation procedures} 

For simulating determinantal point processes, we use the algorithm described in \citet{LMR15} which is a specific case of the simulation algorithm of \citet{Hough2006}. We refer to these references for specific details. The algorithm is implemented in \texttt{spatstat} for the models suggested in \citet{LMR15}, which include Gaussian DPPs. For these models, it is necessary to approximate the kernel because the spectral representation is unknown. In the case of a scaled Ginibre point process, this approximation is however unnecessary for simulating it on a disc because the spectral representation is known. The simulation is still only approximate because the procedure also involves other approximations including approximating the upper bound for rejection sampling chosen in \citet{LMR15} (an approximation which is in fact not necessary for the models they consider since the expression simplifies in those cases). For simulating a scaled Ginibre point process on a window $W$, we thus use the spectral representation on a disc to simulate the process on $b(0,r)\supseteq W$ and thereafter extract the part which is in $W$.

For simulating DPP-Thomas processes on a window $W$, we first simulate the DPP $Y_{\mbox{ext}}$ obtained by restricting $Y$ to an extended window in order to account for boundary effects. Regarding the extension, we decided to use the default setting from the function \texttt{rThomas} in \texttt{spatstat} which simulates a Thomas process. Given the cluster centers $Y_{\mbox{ext}}$ on the extended window, we simulate the clusters of the DPP-Thomas process $X$ independently as finite Poisson processes with intensity functions $\rho_y(x)=\gamma K_{\alpha}(x-y)$ for each $y\in Y_{\mbox{ext}}$. That is, first simulate the number of points $n_y$ in a cluster $X_y$ centered at $y\in Y$ from a Poisson distribution with rate $\gamma$. Then sample the $n_y$ independent points in $X_y$ from the $d$-dimensional normal distribution $N_d(y,\alpha^2I)$. Finally, the simulation of $X$ on $W$ is the part of $\cup_{y\in Y}X_y$ which falls in $W$.

\bibliography{references}

\begin{thebibliography}{}

\bibitem[Baddeley et~al., 2015]{BRT2015}
Baddeley, A., Rubak, E., \& Turner, R. (2015).
\newblock {\em Spatial Point Patterns: Methodology and Applications with {R}}.
\newblock Chapman \& Hall/CRC Press, Boca Raton.

\bibitem[Beraha et~al., 2022]{Beraha22}
Beraha, M., Argiento, R., M{\o}ller, J., \& Guglielmi, A. (2022).
\newblock {MCMC} computations for {B}ayesian mixture models using repulsive
  point processes.
\newblock {\em Journal of Computational and Graphical Statistics}.
\newblock To appear. Available at arXiv:2011.06444.

\bibitem[Biscio \& Lavancier, 2016]{biscio16}
Biscio, C. A.~N. \& Lavancier, F. (2016).
\newblock Quantifying repulsiveness of determinantal point processes.
\newblock {\em Bernoulli}, 22, 2001--2028.

\bibitem[Christoffersen et~al., 2021]{Andreasmfl}
Christoffersen, A.~D., M{\o}ller, J., \& Christensen, H.~S. (2021).
\newblock Modelling columnarity of pyramidal cells in the human cerebral
  cortex.
\newblock {\em Australian and New Zealand Journal of Statistics}, 63, 33--54.

\bibitem[Deng et~al., 2014]{Deng}
Deng, N., Zhou, W., \& Haenggi, M. (2014).
\newblock The {G}inibre point process as a model for wireless networks with
  repulsion.
\newblock {\em IEEE Transactions on Wireless Communications}, 14, 107--121.

\bibitem[Ginibre, 1965]{Ginibre}
Ginibre, J. (1965).
\newblock Statistical ensembles of complex, quaternion, and real matrices.
\newblock {\em Journal of Mathematical Physics}, 6, 440--449.

\bibitem[Guan, 2006]{Guan}
Guan, Y. (2006).
\newblock A composite likelihood approach in fitting spatial point process
  models.
\newblock {\em Journal of American Statistical Association}, 101, 1502--1512.

\bibitem[Hough et~al., 2006]{Hough2006}
Hough, J.~B., Krishnapur, M., Peres, Y., \& Vir{\'a}g, B. (2006).
\newblock Determinantal processes and independence.
\newblock {\em Probability Surveys}, 3, 206--229.

\bibitem[Hough et~al., 2009]{Hough:etal:09}
Hough, J.~B., Krishnapur, M., Peres, Y., \& Vir{\'a}g, B. (2009).
\newblock {\em Zeros of {G}aussian Analytic Functions and Determinantal Point
  Processes}.
\newblock Providence: American Mathematical Society.

\bibitem[Lavancier et~al., 2015]{LMR15}
Lavancier, F., M{\o}ller, J., \& Rubak, E. (2015).
\newblock Determinantal point process models and statistical inference.
\newblock {\em Journal of the Royal Statistical Society: Series B (Statistical
  Methodology)}, 77, 853--877.

\bibitem[Macchi, 1975]{Macchi:75}
Macchi, O. (1975).
\newblock The coincidence approach to stochastic point processes.
\newblock {\em Advances in Applied Probability}, 7, 83--122.

\bibitem[Miyoshi \& Shirai, 2016]{Miyoshi}
Miyoshi, N. \& Shirai, T. (2016).
\newblock Spatial modeling and analysis of cellular networks using the
  {G}inibre point process: {A} tutorial.
\newblock {\em IEICE Transactions on Communications}, 99, 2247--2255.

\bibitem[M{\o}ller, 2003]{moeller:02}
M{\o}ller, J. (2003).
\newblock Shot noise {C}ox processes.
\newblock {\em Advances in Applied Probability}, 35, 614--640.

\bibitem[M{\o}ller \& O'Reilly, 2021]{MReilly21}
M{\o}ller, J. \& O'Reilly, E. (2021).
\newblock Couplings for determinantal point processes and their reduced {P}alm
  distributions with a view to quantifying repulsiveness.
\newblock {\em Journal of Applied Probability}, 58, 469--483.

\bibitem[M{\o}ller \& Torrisi, 2005]{moeller:torrisi:2005}
M{\o}ller, J. \& Torrisi, G.~L. (2005).
\newblock Generalised shot noise {C}ox processes.
\newblock {\em Advances in Applied Probability}, 37, 48--74.

\bibitem[M{\o}ller \& Waagepetersen, 2004]{MW2004}
M{\o}ller, J. \& Waagepetersen, R. (2004).
\newblock {\em Statistical Inference and Simulation for Spatial Point
  Processes}.
\newblock Chapman \& Hall/CRC, Boca Raton, Florida.

\bibitem[M{\o}ller \& Waagepetersen, 2017]{JMRW}
M{\o}ller, J. \& Waagepetersen, R. (2017).
\newblock Some recent developments in statistics for spatial point patterns.
\newblock {\em Annual Review of Statistics and Its Applications}, 4, 317--342.

\bibitem[Mrkvi{\v{c}}ka et~al., 2020]{GET2018}
Mrkvi{\v{c}}ka, T., Myllym{\"a}ki, M., Jilik, M., \& Hahn, U. (2020).
\newblock A one-way {ANOVA} test for functional data with graphical
  interpretation.
\newblock {\em Kybernetika}, 56, 432--458.

\bibitem[Myllym{\"a}ki \& Mrkvi{\v{c}}ka, 2019]{GETinR}
Myllym{\"a}ki, M. \& Mrkvi{\v{c}}ka, T. (2019).
\newblock {GET}: Global envelopes in {R}.
\newblock Available at arXiv:1911.06583.

\bibitem[Myllym{\"a}ki et~al., 2017]{GET2017}
Myllym{\"a}ki, M., Mrkvi{\v{c}}ka, T., Grabarnik, P., Seijo, H., \& Hahn, U.
  (2017).
\newblock Global envelope tests for spatial processes.
\newblock {\em Journal of the Royal Statistical Society: Series B (Statistical
  Methodology)}, 79, 381--404.

\bibitem[{R Core Team}, 2019]{R}
{R Core Team} (2019).
\newblock {\em R: A Language and Environment for Statistical Computing}.
\newblock R Foundation for Statistical Computing, Vienna, Austria.

\bibitem[Ripley, 1976]{ripley:76}
Ripley, B.~D. (1976).
\newblock The second-order analysis of stationary point processes.
\newblock {\em Journal of Applied Probability}, 13, 255--266.

\bibitem[Ripley, 1977]{ripley:77}
Ripley, B.~D. (1977).
\newblock Modelling spatial patterns (with discussion).
\newblock {\em Journal of the Royal Statistical Society: Series B (Statistical
  Methodology)}, 39, 172--212.

\bibitem[Tanaka et~al., 2008]{TanakaEtAl}
Tanaka, U., Ogata, Y., \& Stoyan, D. (2008).
\newblock Parameter estimation and model selection for {N}eyman-{S}cott point
  processes.
\newblock {\em Biometrical Journal}, 50, 43--57.

\bibitem[Thomas, 1949]{Thomas}
Thomas, M. (1949).
\newblock A generalization of {P}oisson's binomial limit in the use in ecology.
\newblock {\em Biometrika}, 36, 18--25.

\bibitem[Van~Lieshout \& Baddeley, 2002]{Lies02}
Van~Lieshout, M. N.~M. \& Baddeley, A.~J. (2002).
\newblock Extrapolating and interpolating spatial patterns.
\newblock In A.~B. Lawson \& D. Denison (Eds.), {\em Spatial Cluster Modelling}
   (pp.\ 61--86). Boca Raton, FL: Chapman and Hall/CRC.

\bibitem[Wickham, 2016]{ggplot}
Wickham, H. (2016).
\newblock {\em ggplot2: Elegant Graphics for Data Analysis}.
\newblock Springer-Verlag New York.

\end{thebibliography}

\end{document}